\documentclass[aps,pre,10pt,twocolumn,floatfix,amsmath]{revtex4-1}

\usepackage{epsfig}

\newcommand{\ep}{\epsilon}

\begin{document}

\title{On the local nature and scaling of chaos in weakly nonlinear
disordered chains}
\author{D. M. Basko}
\affiliation{Universit\'e Grenoble 1/CNRS, LPMMC UMR 5493, 25 rue des
Martyrs, 38042 Grenoble, France}
\begin{abstract}
The dynamics of a disordered nonlinear chain can be either regular
or chaotic with a certain probability. The chaotic behavior is often
associated with the destruction of Anderson localization by the
nonlinearity. In the present work, it is argued that at weak
nonlinearity chaos is nucleated locally on rare resonant segments of
the chain. Based on this picture, the probability of chaos is
evaluated analytically. The same probability is also evaluated by
direct numerical sampling of disorder realizations, and quantitative
agreement between the two results is found.
\end{abstract}

\maketitle

\section{Introduction}

Dynamics of classical disordered nonlinear chains is governed by
an interplay of two fundamental phenomena: Anderson localization
(AL) and chaos. AL, originally introduced as full suppression of
diffusion by interference in random lattices~\cite{Anderson1958}
and used to explain electronic metal-insulator transitions,
turned out to be a universal wave phenomenon and was observed in
such diverse systems as microwaves, light, acoustic waves, and
ultracold atoms~\cite{Lagendijk2009}.
It is most pronounced in one dimension, where all
eigenmodes of a disordered linear system are exponentially
localized by arbitrarily weak disorder~\cite{Gertsenshtein1959}.

By now, AL is well understood for linear systems~\cite{Evers2008}.
In the presence of a nonlinearity, the situation is more
complicated, and many fundamental questions are still open.
For example, it is unclear what happens to an initially localized
wave packet at very long times (see Ref.~\cite{Fishman2012} for a
review). In a linear system with AL, it remains exponentially
localized at all times.
With nonlinearity, the wave packet width was found to increase
as a subdiffusive power of time~$t$ in numerical simulations
\cite{Shepelyansky1993,Pikovsky2008,Flach2009,Bodyfelt2011}.
In contrast, a rigorous argument shows that at long times the
spreading, if any, must be slower than any power of time
\cite{Wang2009}. Analysis of perturbation theory in the
nonlinearity suggests that there is a front propagating as
$~\ln{t}$ beyond which the wave packet is localized
exponentially~\cite{Fishman2009}.
An indication for a slowing down of the power law has also
been seen in the scaling analysis of numerical
results~\cite{Mulansky2011}.
Finally, a possible mechanism of breakdown of subdiffusion
at long times is presented in Ref.~\cite{Michaely2012}.

One of the difficulties in describing nonlinear system is that
their dynamics can be chaotic
\cite{Chirikov1979,Zaslavsky,Lichtenberg}.
For Hamiltonian systems, close to integrable and with a finite
number of the degrees of freedom, the volume of the chaotic
phase space is small as long as the integrability-breaking
perturbation (in our case, the coupling between the oscillators
or the nonlinearity) is weak, as guaranteed by the
Kolmogorov-Arnol'd-Moser (KAM) theorem.

For a random system, the description of chaotic and regular
motion has to be probabilistic.
In the pioneering work~\cite{Frohlich1986}, the existence of
a dense set of regular trajectories was proven for a class of
disordered weakly-nonlinear lattices. It has been argued for
the disordered nonlinear Schr\"odinger chain (DNLS) that the
measure of such set in the phase space is
finite~\cite{Johansson2010}.
This measure, averaged over the disorder, was recently
estimated for DNLS both analytically~\cite{Basko2011} and
numerically~\cite{Pikovsky2011}. In particular, when the
nonlinearity is weak, chaos should appear \emph{locally}.
Namely, for a sufficiently long chain, arbitrarily separated
in two segments, the probability to be on a regular
trajectory is given by the product of the two individual
probabilities for the segments. This naturally follows from
two conditions: (i)~if any of the segments is chaotic, the
whole chain is chaotic, and (ii)~the probabilities for the
two segments are \emph{independent}.
Moreover, in Ref.~\cite{Basko2011} an explicit mechanism
for chaos generation was proposed:
two coinciding resonances for rare combinations of three
oscillators.
Still, the results of Refs.~\cite{Basko2011,Pikovsky2011},
did not agree quantitatively, and the origin of the
discrepancy is currently not understood. Evidence for a
local origin of chaos has also been found in the simulations
of the dynamics of a classical spin chain~\cite{Pal2009}.

In the present work, the probability of chaotic behavior
(appearance of a non-zero Lyapunov exponent) is studied
for another system of coupled nonlinear oscillators [see
Eq.~(\ref{HFSW=}) below], which turns out to be simpler to
analyze than the DNLS. The probability of chaos is calculated
at low energy densities~$\ep$ by two different methods:
(i)~by the analysis of the phase space of an effective
Hamiltonian describing a resonant triple of oscillators, in
the spirit of Ref.~\cite{Basko2011}, and (ii)~by direct
numerical sampling of many disorder realizations, and
counting those with nonzero Lyapunov exponent, analogously to
Ref.~\cite{Pikovsky2011}.

Our main results are the following. (i)~The two calculations
agree quantitatively, including both the leading low-$\ep$
scaling exponent, and the numerical prefactor, thereby
confirming the dominant role of resonant triples in generation
of chaos at low energies.
(ii)~This agreement sets in at unexpectedly small values
of~$\ep$, while at moderately low~$\ep$ the numerical results
fall on an intermediate asymptotics. This intermediate
asymptotics is not controlled by any small parameter and seems
to exist for purely numerical reasons; the microscopic
mechanism responsible for it remains unclear at the moment.

\section{Statement of the problem}

Here we study the model defined by the Hamiltonian
\begin{equation}\begin{split}\label{HFSW=}
&H(\{p_n,q_n\})={H}_0(\{p_n,q_n\})+H_\mathrm{int}(\{p_n,q_n\})=\\
&=\sum_{n=1}^L
\left(\frac{p_n^2}{2m}+\frac{m\omega_n^2q_n^2}{2}\right)
+\frac{g}4\sum_{n=1}^{L-1}\left(q_{n+1}-q_n\right)^4,
\end{split}\end{equation}
where $q_n,p_n$ are the coordinate and momentum of the $n$th
oscillator, and $\omega_n^2$ are independent random variables
uniformly distributed in the interval
\begin{equation}
\frac{W^2}{2}<\omega_n^2<\frac{3W^2}{2},
\end{equation}
$W$ being the disorder strength.
This model belongs to the class of models considered in
Ref.~\cite{Frohlich1986}. It has also been studied in
Refs.~\cite{Mulansky2011,Ivanchenko2011}, however, in the
latter two works focused on spreading of an initially
localized wave packet, rather than on the probability of
chaos.

The model of Eq.~(\ref{HFSW=}) is different from DNLS in two
aspects. First, it has only one conserved quantity (energy)
in contrast to the two (energy and total norm) for DNLS. 
Second, it has only one dimensionless parameter,
\begin{equation}
\ep_n=\frac{g}{m^2W^4}
\left(\frac{p_n^2}{2m}+\frac{m\omega_n^2q_n^2}{2}\right),
\end{equation}
controlling both the nonlinearity and the
coupling between the oscillators. In DNLS they are controlled
separately by two independent dimensionless parameters, and when
both are small, it matters which one is smaller. In the FSW case,
both limits of weak coupling and weak nonlinearity correspond
to $\ep_n\to{0}$, which makes it easier to analyze than DNLS.
From now on, we measure momenta, coordinates, and time in the
units of $\sqrt{m^3W^4/g}$, $\sqrt{mW^2/g}$, and $1/W$,
respectively, which is equivalent to setting $m,W,g=1$.

For a given realization of disorder $\{\omega_n^2\}$ and a given
initial condition $\{p_n,q_n\}$, the system trajectory is either
regular (quasi-periodic) or chaotic. Characterizing each initial
condition by the typical energy~$\ep$ per oscillator, we define
the probability for the initial condition to be chaotic as
\begin{equation}\begin{split}\label{Pfix=}
{P}(\ep,L)
={}&{}\int\Theta(\{p_n,q_n\},\{\omega_n^2\})\times\\
{}&{}\times\prod_{n=1}^L\delta\!
\left(\frac{p_n^2+\omega_n^2q_n^2}{2}-\ep\right)
\frac{dp_ndq_n}{2\pi/\omega_n}\,d\omega_n^2,
\end{split}\end{equation}
where $\Theta(\{p_n,q_n\},\{\omega_n^2\})=1$ if the trajectory
is chaotic and zero otherwise. Eq.~(\ref{Pfix=}) corresponds to
fixing the energies of all oscillators to be~$\ep$, and will be
referred to as the fixed-$\ep$ ensemble. Alternatively, one can
fix the energies only on average, replacing the $\delta$-function
in Eq.~(\ref{Pfix=}) by
$(1/\ep)\exp[-(p_n^2+\omega_n^2q_n^2)/(2\ep)]$, i.~e. a thermal
distribution with temperature~$\ep$ (provided that $\ep\ll{1}$,
which is our main focus)~\cite{Eperosc}. In the
present paper we will work with the fixed-$\ep$ ensemble
which is easier to handle numerically.
The corresponding initial conditions can be represented as
\begin{equation}
p_n=\sqrt{2\ep}\cos\phi_n,\quad
q_n=\frac{\sqrt{2\ep}}{\omega_n}\sin\phi_n,
\end{equation}
where the phases $\phi_n$ are uniformly distributed on the
interval $[0,2\pi]$.

The property of locality, mentioned in the introduction, leads
to the dependence ${P}(\ep,L)=1-e^{-w(\ep)\,L}$ at
sufficiently large~$L$, where the quantity $w(\ep)$, which we
call average chaotic fraction (as in Ref.~\cite{Basko2011}),
does not depend on~$L$.
The main goal of the present work is to establish its asymptotic
behavior $w(\ep\to{0})$, by two methods: (i)~relating it to the
chaotic phase volume of three resonant oscillators, as in
Ref.~\cite{Basko2011}, and (ii)~by direct numerical sampling,
analogously to Ref.~\cite{Pikovsky2011}. The latter method
also provides a check for the locality hypothesis: if indeed
${P}(\ep,L)=1-e^{-w(\ep)\,L}$, the quantity
\[
\frac{1}L\ln\frac{1}{1-{P}(\ep,L)}
\]
does not depend on~$L$, so that all curves for different~$L$
should collapse on a single curve $w(\ep)$.
Thanks to the relative simplicity of the system defined by
Eq.~(\ref{HFSW=}), both calculations can be carried out all
the way to the final result, which turns out to be
\begin{equation}\label{Ae2=}
w(\ep\to{0})=A\ep^2,
\end{equation}
where $A\approx{1}.37\cdot{10}^{3}$ for the fixed-$\ep$ ensemble.

\section{Calculation}

The reduction of the problem to a few-oscillator
configuration~\cite{Basko2011} is based on the two-resonance
picture for weakly non-integrable systems
\cite{Chirikov1979,Zaslavsky,Lichtenberg}. The strongest
resonant term in the non-integrable perturbation of an
integrable system (the so-called guiding resonance) produces
a separatrix in the system phase space. This separatrix is
destroyed by another term in the perturbation, which
creates a thin stochastic layer in the surrounding part
of the phase space. In a disordered system, the main
contribution to the chaotic phase space comes from those
configurations of disorder and from those regions of the
phase space where both perturbation terms are resonant, so
one cannot really distinguish between the one responsible
for the appearance of the separatrix and the one responsible
for its destruction. It is crucial that \emph{two} resonance
conditions should be met simultaneously.

Resonances involving many oscillators are
expected to give a subleading contribution to the chaotic
phase volume, as the corresponding perturbation terms can
be generated in high orders of perturbation theory, thus
resulting in high powers of~$\ep$. The minimal number of
oscillators needed to generate chaos in the model of
Eq.~(\ref{HFSW=}) is two~\cite{chaosDNLS}.
However, even if $\omega_1\approx\omega_2$, the frequency of
the separatrix-destroying perturbation, $\omega_1+\omega_2$,
cannot be small for the chosen disorder distribution,
$1/2<\omega_n^2<3/2$, which leads to an exponential
suppression of the chaotic phase volume
\cite{Chirikov1979,Zaslavsky,Lichtenberg}.
In fact, for the particular case of Eq.~(\ref{HFSW=}), the
situation is even worse:
no separatrix exists in the phase space of the slow motion
(see Appendix~\ref{app:couple}). Thus, the
dominant contribution to the chaotic phase volume should
come from triples of oscillators with all three frequencies
close to each other. The oscillators should also be
neighboring each other in space, since coupling distant
oscillators requires high orders of perturbation theory and
results in a high power of~$\ep$.
Thus, $w(\ep)$ is essentially determined by the probability
(per unit length along the chain) to find three neighboring
oscillators whose frequencies differ by $\sim\ep$. This
fixes the power $w(\ep)\propto\ep^2$.

To put this argument on a quantitative basis, we assume
$\omega_1\approx\omega_2\approx\omega_3$, and average
Eq.~(\ref{HFSW=}) over fast oscillations (see
Appendix~\ref{app:triple} for details). This gives
the effective Hamiltonian of the resonant triple:
\begin{equation}\begin{split}\label{Htriple=}
H_\mathrm{tr}={}&{}
\Omega|\Psi_1|^2-(\Omega+\Omega')|\Psi_2|^2+\Omega'|\Psi_3|^2+\\
{}&{}+\frac{1}{2}
\left(|\Psi_1|^4+|\Psi_3|^4+|\Psi_1-\Psi_2|^4+|\Psi_2-\Psi_3|^4\right),
\end{split}\end{equation}
written in terms of complex canonical variables
$i\Psi_n^*,\Psi_n$, related to $p_n,q_n$ as
\begin{equation}
\Psi_n=\frac{e^{i\bar\omega{t}}}{\sqrt{I}}
\left(\frac{p_n}{\sqrt{2\omega_n}}-i\sqrt{2\omega_n}q_n\right),\quad
I=\sum_{n=1}^3\frac{p_n^2+\omega_n^2q_n^2}{2\omega_n}.
\end{equation}
where $\bar\omega=(\omega_1+\omega_2+\omega_3)/3$.
The rescaling of $\Psi$'s by $\sqrt{I}$ restricts them
to the unit sphere, $|\Psi_1|^2+|\Psi_2|^2+|\Psi_3|^2=1$,
which is invariant under the dynamics generated by the
Hamiltonian in Eq.~(\ref{Htriple=}). The rescaled
frequency mismatches $\Omega,\Omega'$ are defined as
\begin{equation}\label{Omega=}
\Omega=\frac{2\omega_1-\omega_2-\omega_3}{(9/4)(I/\bar\omega^2)},
\quad
\Omega'=\frac{2\omega_3-\omega_2-\omega_1}{(9/4)(I/\bar\omega^2)},
\end{equation}

For the fixed-$\ep$ ensemble, Eq.~(\ref{Pfix=}), we have
$I=3\ep/\bar\omega+O(\ep^2)$, and the initial condition
should be chosen in the form $\Psi_1=e^{-i\varphi}/\sqrt{3}$,
$\Psi_2=e^{i\varphi+i\varphi'}/\sqrt{3}$
$\Psi_3=e^{-i\varphi'}/\sqrt{3}$, where $0<\varphi,\varphi'<2\pi$
are random phases (the global phase drops out of the result).
Then, defining $\Theta(\varphi,\varphi';\Omega,\Omega')$ to be~1
if the corresponding trajectory is chaotic for given values
of $\varphi,\varphi',\Omega,\Omega'$ and 0~otherwise, we obtain
from Eq.~(\ref{Pfix=}):
\begin{equation}\label{wep=}
w(\ep)=81\ep^2\int\limits_{-\infty}^\infty d\Omega\,d\Omega'
\int\limits_0^{2\pi}
\frac{d\varphi}{2\pi}\frac{d\varphi'}{2\pi}\,
\Theta(\varphi,\varphi';\Omega,\Omega').
\end{equation}
Since neither the Hamiltonian [Eq.~(\ref{Htriple=})] nor the
initial condition contain any small parameters, the integral
over~$\Omega,\Omega'$ in Eq.~(\ref{wep=}) is dominated by
$|\Omega|,|\Omega'|\sim{1}$, while for larger mismatches the
chaotic regions quickly shrink. Thus, the limits of the
$\Omega,\Omega'$-integration (which are $\sim{1}/\ep$) have
been extented to infinity. The factor $\ep^2$ in front of
the integral appears because
$\Omega,\Omega'\propto{1}/I\propto{1}/\ep$,
see Eq.~(\ref{Omega=}). The integral is evaluated numerically
to be $16.9\pm{0.2}$, leading to $w(\ep\to{0})=A\ep^2$ with
$A=1.37\cdot{10}^{3}$, as stated in Eq.~(\ref{Ae2=}).

The key element of the numerical procedure is the criterion
which enables one to distinguish between regular and chaotic
motion. Formally, one studies the largest Lyapunov
exponent~$\sigma$, characterizing the mean exponential rate
of divergence of two initially close trajectories,
\begin{equation}
\sigma=\lim\limits_{t\to\infty}\frac{\Lambda(t)}{t},\quad
\Lambda(t)=\ln\frac{d(t)}{d(0)},
\end{equation}
where $d(t)$ is the distance between points belonging
to the two trajectories~\cite{Lichtenberg}. If $\sigma=0$,
the motion is regular, if $\sigma>0$, it is chaotic.
In practice, one may integrate the equations of motion
for a sufficiently long time~$T$, and consider $\sigma=0$
if $\Lambda(T)/T<(\mbox{a few times})(1/T)$
\cite{Pikovsky2011,Mulansky2011a}.
Here we use a slightly different criterion, based on the
behavior of $\Lambda(t)$ in the whole integration interval
$0<t<T$. Namely, we check how well $\Lambda(t)$ can be
approximated by a logarithmic function, as described in
Appendix~\ref{app:numerics}.

\begin{figure}
\begin{center}
\includegraphics[width=8cm]{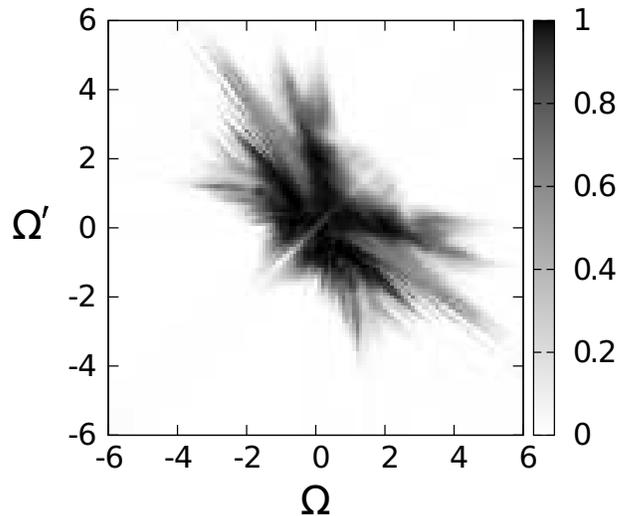}
\end{center}
\caption{\label{fig:fraction}
A grayscale plot of the integral
$\int_0^{2\pi}(d\varphi/2\pi)(d\varphi'/2\pi)\,
\Theta(\varphi,\varphi';\Omega,\Omega')$ as a function
of $\Omega,\Omega'$.
}
\end{figure}

In Fig.~\ref{fig:fraction}, the $\varphi,\varphi'$-integral
in Eq.~(\ref{wep=}) is plotted as a function
of $\Omega,\Omega'$. The fact that the chaotic region is
confined in all directions, represents a numerical proof
of the statement that chaos arises mostly in the regions
where \emph{two} resonant conditions are satisfied
simultaneously.

\begin{figure}
\begin{center}
\includegraphics[width=8cm]{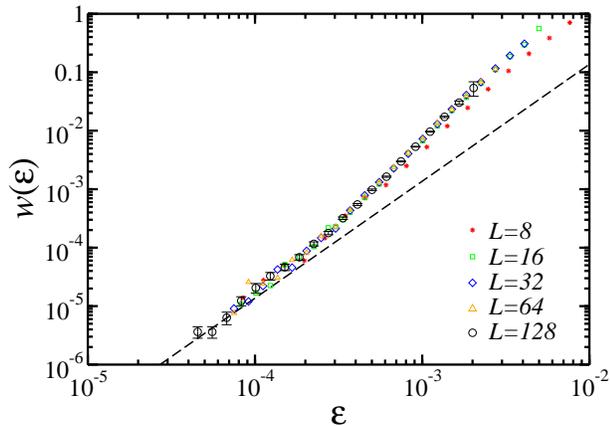}
\end{center}
\caption{\label{fig:wlog}(color online)
$(1/L)|\ln(1-{P}_L(\ep))|$ versus~$\ep$ for the
fixed-density excitation and different values of
$L=8,16,32,64,128$ (symbols). The error bars for
$L=128$ correspond to the relative error $1/\sqrt{N}$,
where $N$ is the absolute number of detected events.
The dashed line represents the dependence
$w(\ep)=(1.37\cdot{10}^{3})\,\ep^2$,
obtained from the resonant-triple calculation.}
\end{figure}

Using the same algorithm, $w(\ep)$ was also calculated by
direct numerical sampling of disorder realizations on chains
of lengths $L=8,16,32,64,128$.
For $\ep>{10}^{-4}$, the integration time $T=10^6$ and
averaging over $10^4$ realizations was performed for each~$L$;
for smaller $\ep<10^{-4}$, when the probability ${P}_L(\ep)$
is also very small, more than $4\cdot{10}^4$ realizations were
required to accumulate at least $N=20$ events for each data
point. Then $1/\sqrt{N}$ was assumed to give the relative
uncertainty of the obtained value $w(\ep)$. 
Also, for $\ep<10^{-4}$, the integration time had to be
increased to $T=3\cdot{10}^6$ in order to reliably distinguish
between regular and chaotic dynamics (see also
Appendix~\ref{app:numerics}). 

The numerical results, as well as the dependence
$w(\ep)=(1.37\cdot{10}^{3})\,\ep^2$, obtained from
Eq.~(\ref{wep=}), are shown in Fig.~\ref{fig:wlog} by the
symbols and the dashed line, respectively. Fig.~\ref{fig:wlog}
represents the main result of the present work.
Starting from $L=16$, the collapse of the numerical data is
very good, which numerically proves the local origin of chaos.
At $\ep<3\cdot{10}^{-4}$ the numerical data fall on the dashed
line. At larger~$\ep$ the data collapses on some intermediate
asymptotics for~$w(\ep)$, which can be approximated by a power
law $w(\ep)=B\ep^\beta$ with the exponent $\beta=2.85\pm{0}.1$
and a surprisingly large prefactor $B=(1-5)\cdot{10}^6$.
The precise mechanism responsible for this intermediate
asymptotics is not clear at the moment. This mechanism is
likely to involve a larger number of oscillators (more than
three), since (i)~it is in this region of $\ep$ that the
symbols for short chains ($L=8$) exhibit a systematic
deviation from those for long chains in Fig.~\ref{fig:wlog},
in the direction of suppression of the intermediate
asymptotics; (ii)~for a single-site initial excitation (see
Fig.~\ref{fig:SingleSite} below) the intermediate
asymptotics is completely absent, indicating that it requires
a certain initial spread.

\section{Relation to wave packet spreading}

In this context, one should mention the result reported in
Ref.~\cite{Ivanchenko2011}, where the probability of
subdiffusive spreading of an initially localized wave packet
was studied for exactly the same model, Eq.~(\ref{HFSW=}).
Namely, it was argued that an initially localized wave packet
either stays localized forever or spreads indefinitely, and
the probability of spreading was estimated to be proportional
to~$\ep$ at small~$\ep$. In the present work, the probability
of chaos, was shown to scale as $\ep^2$. Since the latter is
smaller than the former, the combination of the two results
would suggest that the wave packet spreading does not require
chaos, which is quite counter-intuitive.
However, in the author's opinion, it is more plausible that
the probability of spreading was strongly overestimated in
Ref.~\cite{Ivanchenko2011}, as discussed below.

\begin{figure}
\begin{center}
\vspace{0.5cm}
\includegraphics[width=8cm]{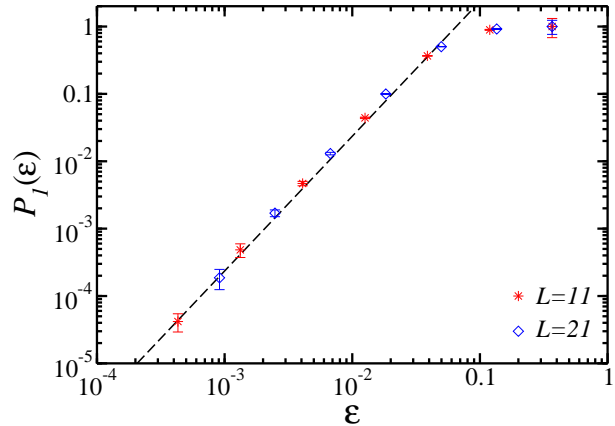}
\vspace{-0.5cm}
\end{center}
\caption{\label{fig:SingleSite}(color online)
The probability of chaos versus~$\ep$ for a single-site
excitation and two values of $L=11,21$ (symbols). The dashed
line represents the dependence $235\,\ep^2$, obtained from the
resonant-triple calculation of Appendix~\ref{app:SingleSite}.}
\end{figure}

First, let us exclude the trivial possibility that the
difference between the two results is simply due to the
fact that initial conditions studied in
Ref.~\cite{Ivanchenko2011} correspond to excitation
localized on a single site, while a finite density was
assumed in the present work. Indeed, for a finite
density the trajectory is chaotic when a resonant
triple occurs anywhere on the chain, while for a single-site
excitation the condition is simply that the resonant triple
includes the excited site. This affects the scaling with~$L$,
but not with~$\ep$: the probability of chaos for a single-site
excitation scales as $P_1(\ep)=A_1\ep^2$ and is independent
of~$L$. The resonant-triple calculation gives
$A_1\approx{235}$ (see Appendix~\ref{app:SingleSite}), which
is also confirmed by the direct numerical sampling, as shown
in Fig.~\ref{fig:SingleSite}. (Note also the direct crossover
from 1 to $A_1\ep^2$ without any intermediate asymptotics).

The analytical arguments of
Ref.~\cite{Ivanchenko2011} are based on the assumption that a
\emph{single} resonance is sufficient for spreading.
This immediately gives the probability scaling~$\propto\ep$.
However, it is not clear why a single resonance should lead
to unlimited spreading; indeed, an isolated nonlinear
resonance is known to produce just periodic
oscillations~\cite{Chirikov1979,Zaslavsky}.

The numerical procedure of Ref.~\cite{Ivanchenko2011} was
based on the study of the participation number,
\begin{equation}\label{Pdef=}
\mathcal{P}=\left(\sum_{n=1}^LE_n\right)^2
\left(\sum_{n=1}^LE_n^2\right)^{-1},
\end{equation}
where the on-site energy corresponding to the model of
Eq.~(\ref{HFSW=}) can be defined as
\begin{equation}
E_n=\frac{p_n^2}{2m}+\frac{m\omega_n^2q_n^2}{2}
+\frac{g}8\left[(q_n-q_{n-1})^4+(q_n-q_{n+1})^4\right],
\end{equation}
for all $n$ except $n=1$ and $n=L$, where only one nonlinear
term corresponding to the unique neighbor should be taken.
For a perfectly thermalized chain $\mathcal{P}=L/2$, and for a
state perfectly localized on a single site $\mathcal{P}=1$.
In Ref.~\cite{Ivanchenko2011}, a trajectory was counted as
spreading if $\mathcal{P}$ exceeded an arbitrarily chosen 
threshold of 1.2 by the time $t=10^9$. Thus, the main
assumption behind the numerics was that if a trajectory has
overcome the limit $\mathcal{P}=1.2$ at $t=10^9$, it will
spread forever. In the following, it is argued that this is
unlikely to be true for the majority of the trajectories.

\begin{figure}
\begin{center}
\vspace{0.5cm}
\includegraphics[width=8cm]{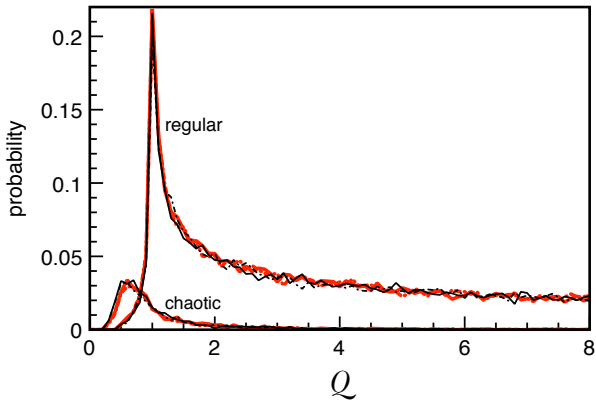}
\vspace{-0.5cm}
\end{center}
\caption{\label{fig:ipr} (color online)
Partial contributions to the probability distribution of
$\mathcal{Q}=1/(\mathcal{P}-1)$ [see Eq.~(\ref{Pdef=})]
from regular and chaotic trajectories (upper and lower
curves, respectively) at times $t=10^5,10^6$ [thick red
(gray) and thin black curves, respectively] for two chain
lengths $L=11,21$ (dotted and solid curves, respectively)
for $\ep=0.01$.}
\end{figure}

Let us analyze the
probability distribution of $\mathcal{Q}=1/(\mathcal{P}-1)$,
which is a more convenient quantity to analyze at small~$\ep$,
when the spreading trajectories correspond to
$\mathcal{Q}\sim{1}$, while the overwhelming majority of
strongly localized solutions form the broad large-$\mathcal{Q}$
tail.
In Fig.~\ref{fig:ipr} we separately plot the two contributions
to the distribution of $\mathcal{Q}$ from regular and chaotic
trajectories at $t=10^5,10^6$ for $L=11,21$. The value of
$\ep=0.01$ is sufficiently small to be in the asymptotic
regime, as seen from Fig.~\ref{fig:SingleSite}, and more than
50000 disorder realizations were used for each of the four
pairs of curves.
The difference between $L=11$ and $L=21$ curves is not
detectable within the numerical resolution, while a slight
difference between $t=10^5$ and $t=10^6$ curves can be seen
on the low-$\mathcal{Q}$ side of the chaotic curves, which
indicates some spreading of the wave packet.
The fact that a large part of the probability distribution
at $\mathcal{Q}<5$ (corresponding to $\mathcal{P}>1.2$)
belongs to the large-$\mathcal{Q}$ tail due to localized
trajectories, suggests that many trajectories counted as
spreading in Ref.~\cite{Ivanchenko2011}, in fact were not.

Another argument, not relying on numerical integration over
long times, can be given by considering a regular solution,
predominantly localized on site~$n$. The amplitude on the
neighboring sites $n-1,n+1$ can be found by perturbation
theory,
\begin{subequations}
\begin{eqnarray}
&&q_n=\frac{\sqrt{2\ep}}{\omega_n}\,\cos\omega_nt+O(\ep^{3/2}),\\
&&q_{n\pm{1}}=\frac{\ep^{3/2}}{\sqrt{2}\omega_n^3}
\left(\frac{3\cos\omega_nt}{\omega_{n\pm{1}}^2-\omega_n^2}
+\frac{\cos{3}\omega_nt}{\omega_{n\pm{1}}^2-9\omega_n^2}\right)
+O(\ep^{5/2}),\nonumber\\
\end{eqnarray}
\end{subequations}
which gives the participation number
\begin{equation}\begin{split}\label{perturb=}
\mathcal{P}\approx{}&{}{1}
+2\,\frac{\ep_{n-1}+\ep_{n+1}}{\ep_n}\approx\\
\approx{}&{}1+\frac{9\ep^2}{4\omega_n^6}\sum_\pm\left[
\frac{\omega_{n\pm{1}}^2+\omega_n^2}{(\omega_{n\pm{1}}^2-\omega_n^2)^2}
+\frac{\omega_{n\pm{1}}^2/9+\omega_n^2}%
{(\omega_{n\pm{1}}^2-9\omega_n^2)^2}\right].
\end{split}\end{equation}
Here we omitted oscillating terms, which at large $t\gg{1}$
quickly vanish upon disorder averaging), as well as terms
which do not contain potentially small denominators and thus
are bounded by $O(\ep)$ [such as $3\ep/(4\omega_n^4)$,
originating from the $(q_{n\pm{1}}-q_n)^4$ contribution to
$E_{n\pm{1}}$].

From Eq.~(\ref{perturb=}),
one can find the probability of $\mathcal{P}>1.2$
at small $\ep$ by noting that $\mathcal{P}>1.2$ may occur
when either $\omega_{n+1}$ or $\omega_{n-1}$ is close to
$\omega_n$. The probability of both occurring simultaneously
can be neglected, as well as the probability of
$\omega_{n\pm{1}}$ being close to $3\omega_n$, as these
probabilities are $\propto\ep^2$. Then, the probability of
$\mathcal{P}>1.2$ is given by
\begin{equation}\begin{split}\label{proptoep=}
\mathop{\mathrm{Pr}}\left\{\mathcal{P}>1.2\right\}
\approx{}&{}\sum_\pm
\mathop{\mathrm{Pr}}\left\{\frac{9\ep^2}{4\omega_n^6}\,
\frac{\omega_{n\pm{1}}^2+\omega_n^2}%
{(\omega_{n\pm{1}}^2-\omega_n^2)^2}>0.2\right\}\approx\\
\approx{}&{}2\mathop{\mathrm{Pr}}\left\{
|\omega_{n+1}^2-\omega_n^2|<\frac{1}{\sqrt{0.2}}\,
\frac{3\ep}{\sqrt{2}\omega_n^2}\right\}\approx\\
\approx{}&{}\frac{3\sqrt{8}\,\ep}{\sqrt{0.2}}
\int\limits_{1/2}^{3/2}\frac{d\omega_n^2}{\omega_n^2}
\approx{20}.8\,\ep.
\end{split}\end{equation}
The coefficient in front of~$\ep$ is very close to that
obtained numerically for the probability of spreading
(see inset of Fig.~3 of Ref.~\cite{Ivanchenko2011}).

The fact that Eq.~(\ref{proptoep=}), derived under the
assumption of validity of perturbation theory at
arbitrarily long times (i.~e., that the trajectory is
localized), gives practically the same result as the
numerics of Ref.~\cite{Ivanchenko2011}, means that almost
all trajectories counted as spreading in
Ref.~\cite{Ivanchenko2011}, were in fact localized. Thus,
the question of whether the probability of unlimited wave
packet spreading scales as $\propto\ep$, or $\propto\ep^2$,
or is identically zero, remains open.

\section{Conclusions}

In this work, the probability for a long chain of oscillators,
defined by its Hamiltonian, Eq.~(\ref{HFSW=}), to be on a
chaotic trajectory, was calculated by analyzing the chaotic
phase space of rare resonant configurations of three
oscillators. The result, $1-e^{-A\ep^2L}$, agrees with the
direct numerical evaluation of Lyapunov exponents for many
disorder realizations and initial conditions at sufficiently
small values of $\ep$, the energy per oscillator. While the
coefficient~$A$ depends on the specific ensemble used for the
initial conditions, the power
law $\ep^2$ is determined by the mere fact that chaotic phase
volume is dominated by the regions of the phase space where
\emph{two} resonant conditions for three neighboring oscillators
are satisfied simultaneously.
This fact was numerically verified for the three-oscillator
configuration, see Fig.~\ref{fig:fraction}.

The following main conclusions can be drawn from the results
of present work.
(i)~The local nature of chaos, established earlier for the
discrete nonlinear Schr\"odinger equation with disorder
(DNLS) in Refs.~\cite{Basko2011,Pikovsky2011} for weak
nonlinearity, is further confirmed here for a different model.
Namely, the probability of regular dynamics decays with
length~$L$ as $e^{-wL}$, so $w$~can be called the probability
per unit length for the chaos to occur.
(ii)~Two different ways to calculate this probability, namely,
from the analysis of the phase space of a resonant triple, and
by direct numerical sampling of disorder realizations, give the
same result, but this agreement sets in at unexpectedly small
values of~$\ep\sim{3}\times{10}^{-4}$. At larger~$\ep$, an
intermediate asymptotics is seen in the numerical results. This
fact may be relevant for understanding the disagreement between
the results of Refs.~\cite{Basko2011} and~\cite{Pikovsky2011}
for the chaotic probability in the DNLS; indeed, one cannot
exclude existence of similar intermediate regimes in DNLS.
However, the results of the present work were obtained for a
quite different model with a different number of independent
parameters, so no quantitative comparison can be made with the
DNLS, and the latter should be studied further.

\section{Acknowledgements}
The author is grateful to M. Ivanchenko, T. Laptyeva, and S. Flach
for clarifying discussions of Ref.~\cite{Ivanchenko2011},
and to S. Fishman for a critical reading of the manuscript.

\appendix

\section{Few-oscillator configurations}

It is convenient to pass to the canonical action-angle
variables, $(p_n,q_n)\to(I_n,\phi_n)$:
\begin{equation}
p_n=\sqrt{2\omega_nI_n}\,\cos\phi_n,\quad
q_n=\sqrt{\frac{2I_n}{\omega_n}}\,\sin\phi_n.
\end{equation}
Hamiltonian (\ref{HFSW=}), averaged over the fast oscillations,
(that is, with terms depending on $\phi_n+\phi_{n+1}$ omitted)
takes the following form:
\begin{widetext}\begin{equation}\begin{split}\label{HRWA=}
H={}&{}\sum_{n=1}^L\omega_nI_n
+\frac{3}{8}
\sum_{n=1}^{L-1}\left[\frac{I_n}{\omega_n}
+\frac{I_{n+1}}{\omega_{n+1}}
-2\sqrt{\frac{I_nI_{n+1}}{\omega_n\omega_{n+1}}}
\cos(\phi_n-\phi_{n+1})\right]^2.
\end{split}\end{equation}\end{widetext}
It should be noted that the part of the Hamiltonian
corresponding to two or three oscillators (say, $n,n+1,n+2$)
far from the ends of the chain is different from the Hamiltonian
given by Eq.~(\ref{HRWA=}) for $L=2$ or $L=3$. Indeed, in the
former case the Hamiltonian of the $n$th oscillator contains the
term $(3/4)I_n^2/\omega_n^2$, coming from coupling to the two
neighbors $n-1$ and $n+1$, while in the latter case the $n=1$
oscillator has only one neighbor, so the nonlinear term is
twice smaller, $(3/8)I_1^2/\omega_1^2$.

\subsection{Two oscillators}\label{app:couple}

Let $\omega_n\approx\omega_{n+1}$ be strongly different
from $\omega_{n-1},\omega_{n+2}$.
We perform a canonical change of variables of the pair
\begin{equation}\begin{split}
&I=I_n+I_{n+1},\quad\phi=\frac{\phi_n+\phi_{n+1}}{2},\\
&J=\frac{I_n-I_{n+1}}2,\quad\varphi=\phi_n-\phi_{n+1},
\end{split}\end{equation}
and further denote  $J=(I/2)\cos\vartheta$. Then the
Hamiltonian of the pair is given by
\begin{equation}\begin{split}
H={}&{}\frac{\omega_1'+\omega_2'}{2}\,I
+(\omega_1'-\omega_2')(I/2)\cos\vartheta{}+{}\\
{}&{}+\frac{3}4\,\frac{I^2}4\left(\frac{\omega_1+\omega_2}{\omega_1\omega_2}
-\frac{\omega_1-\omega_2}{\omega_1\omega_2}\,\cos\vartheta\right)^2{}+{}\\
{}&{}+\frac{3}2\,\frac{(I^2/4)\sin\vartheta}{(\omega_1\omega_2)^{3/2}}
\left[\omega_1+\omega_2-(\omega_1-\omega_2)\cos\vartheta\right]\cos\varphi{}+{}\\
{}&{}+\frac{3}4\,\frac{I^2/4}{\omega_1\omega_2}\,
\sin^2\vartheta\cos{2}\varphi,
\end{split}\end{equation}
where $\omega_1'=\omega_n+4I_{n-1}/(\omega_n\omega_{n-1})$,
$\omega_2'=\omega_{n+1}+4I_{n+2}/(\omega_{n+1}\omega_{n+2})$,
We can assume $I_{n-1},I_{n+2}$ to be constant, as their
changes due to weak non-resonant perturbation are small.
This Hamiltonian can be approximately rewritten as
\begin{subequations}\begin{eqnarray}
&&H\approx\frac{3}{16}\frac{I^2}{\bar\omega^2}
\left(\Omega\cos\vartheta+4\sin\vartheta\cos\varphi
+\sin^2\vartheta\cos{2}\varphi\right)+\nonumber\\
&&\qquad\qquad{}+\mathrm{const},\\
&&\Omega=\frac{8\bar\omega^2(\omega_1'-\omega_2')}{3I},\quad
\bar\omega=\frac{\omega_1+\omega_2}{2}.
\end{eqnarray}\end{subequations}
This Hamiltonian has only two elliptic stationary points in
the phase space $(\vartheta,\varphi)$ at any value of the
rescaled mismatch~$\Omega$. 
Hence, its phase space does not contain a separatrix at all,
so we do not expect any chaos at small~$\ep$.

\subsection{Three oscillators}
\label{app:triple}

As we are interested in the properties of a resonant triple
$n,n+1,n+2$
far from the ends of the chain, let us denote the three
frequencies by
\begin{equation}\begin{split}\label{omegaprime=}
&\omega^\prime_1=\omega_n+\frac{4I_{n-1}}{\omega_n\omega_{n-1}},\\
&\omega^\prime_2=\omega_{n+1},\\
&\omega^\prime_3=\omega_{n+2}+\frac{4I_{n+3}}{\omega_{n+2}\omega_{n+3}},
\end{split}\end{equation}
and assume $I_{n-1}$ and $I_{n+3}$ to be constant, since
their variation is small in the parameter~$\ep\ll{1}$.
Moreover, we can also neglect the modification in the
probability distributions of $\omega^\prime_1,\omega^\prime_3$
with respect to that of $\omega_n$, as the difference is
again $\sim\ep$. 
The most important contribution to the chaotic fraction
will come from the region where the differences
$\omega^\prime_1-\omega^\prime_2$,
$\omega^\prime_2-\omega^\prime_3$,
are small compared to the frequencies themselves.
Having this in mind, let us further denote
\begin{equation}
\bar\omega=\frac{\omega^\prime_1+\omega^\prime_2+\omega^\prime_3}{3}.
\end{equation}

Introducing the complex canonical coordinates
$\psi_1=\sqrt{I_n}\,e^{-i\phi_n}$,
$\psi_2=\sqrt{I_{n+1}}\,e^{-i\phi_{n+1}}$,
$\psi_3=\sqrt{I_{n+2}}\,e^{-i\phi_{n+2}}$,
whose conjugate momenta are $i\psi_1^*,i\psi_2^*,i\psi_3^*$,
respectively, we can write the Hamiltonian of the triple as
\begin{equation}\begin{split}\label{HtripleA=}
H={}&\sum_{i=1}^3\omega_i'\psi_i^*\psi_i
+\frac{3}{8}\,\frac{\psi_1^*\psi_1^*\psi_1\psi_1}{(\omega_1')^2}
+\frac{3}{8}\,\frac{\psi_3^*\psi_3^*\psi_3\psi_3}{(\omega_3')^2}\\
{}&+\frac{3}8\sum_{i=1}^2
\left(\frac{\psi_i^*}{\sqrt{\omega_i'}}
-\frac{\psi_{i+1}^*}{\sqrt{\omega_{i+1}'}}\right)^2
\left(\frac{\psi_i}{\sqrt{\omega_i'}}
-\frac{\psi_{i+1}}{\sqrt{\omega_{i+1}'}}\right)^2.
\end{split}\end{equation}
In the nonlinear terms (those of the fourth order in
$\psi_i,\psi_i^*$) it is sufficient to replace
$\omega^\prime_{1,2,3}\to\bar\omega$, to the leading
order in~$\ep$.
The total norm,
\begin{equation}
I=\psi_1^*\psi_1+\psi_2^*\psi_2+\psi_3^*\psi_3,
\end{equation}
is conserved for the Hamiltonian~(\ref{HtripleA=}), and the
corresponding conjugate phase,
\begin{equation}
\phi=\frac{\phi_{n}+\phi_{n+1}+\phi_{n+2}}{3},
\end{equation}
even though depends on time in some complicated way,
does not contribute to anything.
In the fixed-$\ep$ ensemble, we simply have
$I=3\ep/\bar\omega+O(\ep^2)$.
It is convenient to pass to rescaled variables
\begin{equation}\begin{split}
&\Psi_i=\frac{e^{i\bar\omega{t}}}{\sqrt{I}}\,\psi_i,\quad i=1,2,3,\\
&\Omega=\frac{4}3\,\frac{\bar\omega^2}{I}\,
\frac{2\omega_1'-\omega_2'-\omega_3'}3,\\
&\Omega'=\frac{4}3\,\frac{\bar\omega^2}{I}\,
\frac{2\omega_3'-\omega_2'-\omega_1'}{3}.
\end{split}\end{equation}
Then $|\Psi_1|^2+|\Psi_2|^2+|\Psi_3|^2=1$.
The probability density for frequencies is
\begin{equation}\begin{split}
\nu(\bar\omega)=&{}\lim\limits_{\Omega,\Omega'\to{0}}
\int\limits_{1/2}^{3/2}dx\,dy\,dz\,
\delta\!
\left(\frac{\sqrt{x}+\sqrt{y}+\sqrt{z}}3-\bar\omega\right)\\
&{}\times
\delta\!\left(\frac{2\sqrt{x}-\sqrt{y}-\sqrt{z}}%
{(9/4)(I/\bar\omega^2)}-\Omega\right)\\
&{}\times
\delta\!\left(\frac{2\sqrt{z}-\sqrt{y}-\sqrt{x}}%
{(9/4)(I/\bar\omega^2)}-\Omega'\right)
\\
={}&\frac{27}2\,\frac{I^2}{\bar\omega}+O(\ep^2).
\end{split}\end{equation}
The rescaled Hamiltonian is given by
\begin{equation}\begin{split}
\frac{H-\bar\omega{I}}{(3/4)I^2\bar\omega^{-2}}={}&{}
\Omega|\Psi_1|^2-(\Omega+\Omega')|\Psi_2|^2+\Omega'|\Psi_3|^2+\\
{}&{}+\frac{1}{2}\,|\Psi_1|^4+\frac{1}{2}\,|\Psi_3|^4+\\
{}&+\frac{1}{2}\left|\Psi_1-\Psi_2\right|^4+
\frac{1}{2}\left|\Psi_2-\Psi_3\right|^4.
\end{split}\end{equation}
For the thermal distribution the initial condition should be
taken in the form
\begin{equation}\begin{split}
&\Psi_1=\sqrt{1-x}\,e^{i(-2\varphi_1+\varphi_2)/3},\\
&\Psi_2=\sqrt{x-y}\,e^{i(\varphi_1+\varphi_2)/3},\\
&\Psi_3=\sqrt{y}\,e^{i(\varphi_1-2\varphi_2)/3}.
\end{split}\end{equation}
Then the chaotic fraction is given by
\begin{equation}\begin{split}
w(\ep)={}&\int\limits_0^\infty
e^{-\bar\omega{I}/\ep}\,\frac{I^2\,dI}{(\ep/\bar\omega)^3}\,
\int\limits_{\sqrt{1/2}}^{\sqrt{3/2}}
\frac{27}{2}\,\frac{I^2}{\bar\omega}\,d\bar\omega
\int\limits_{-\infty}^\infty d\Omega\,d\Omega'
\times{}\\
{}&\times\int\limits_0^{2\pi}
\frac{d\varphi_1}{2\pi}\frac{d\varphi_2}{2\pi}
\int\limits_0^1dx\int\limits_0^xdy\,
\Theta(x,y;\varphi_1,\varphi_2;\Omega,\Omega').
\end{split}\end{equation}
The first two integrals amount to $216\,\ep^2$.
The fixed-$\ep$ ensemble corresponds to fixed
$x=2/3$, $y=1/3$, so the chaotic fraction is given by
\begin{equation}\begin{split}
w(\ep)={}&\int\limits_{\sqrt{1/2}}^{\sqrt{3/2}}
\frac{27}{2}\,\frac{(3\ep)^2}{\bar\omega^3}\,d\bar\omega
\int\limits_{-\infty}^\infty d\Omega\,d\Omega'
\times{}\\
{}&\times\int\limits_0^{2\pi}
\frac{d\varphi_1}{2\pi}\frac{d\varphi_2}{2\pi}\,
\Theta(2/3,1/3;\varphi_1,\varphi_2;\Omega,\Omega').
\end{split}\end{equation}
The first integral is equal to $81\,\ep^2$, and
the integral over $\Omega,\Omega'$ to $16.9\pm{0}.2$.

\begin{figure}
\begin{center}
\includegraphics[width=8cm]{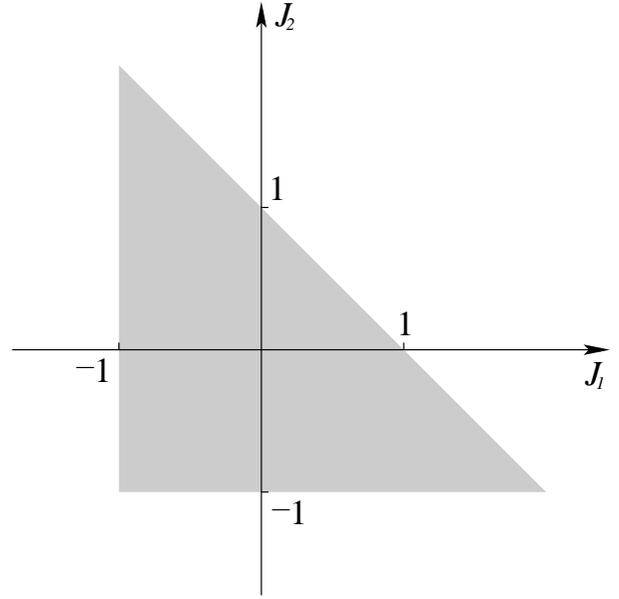}
\end{center}
\caption{\label{fig:triangle12}
The allowed region of $J_1,J_2$, shown by the shaded area.
}
\end{figure}

One can separate the conserved total norm and write
the explicit Hamiltonian for the two remaining degrees
of freedom. Let us label the actions and phases
of the oscillators $n,n+1,n+2$ by $i=1,2,3$, respectively.
Then the corresponding canonical transformation is
determined by
\begin{equation}\begin{split}
&I=I_1+I_2+I_3,\quad
\phi=\frac{\phi_1+\phi_2+\phi_3}{3},\\
&\frac{I}{3}\,{J}_1=\frac{2I_1-I_2-I_3}{3},\quad
\varphi_1=\phi_1-\phi_2,\\
&\frac{I}{3}\,{J}_2=\frac{I_1+I_2-2I_3}{3},\quad
\varphi_2=\phi_2-\phi_3.
\end{split}\end{equation}
To ensure $I_1,I_2,I_3>0$, the variables
$J_1,J_2$ should lie inside the triangle, shown in
Fig.~\ref{fig:triangle12}. The inverse transformation is
\begin{equation}\begin{split}
&\psi_1=\sqrt{(1+J_1)I/3}\,e^{i(-2\varphi_1+\varphi_2)/3-i\phi},\\
&\psi_2=\sqrt{(1-J_1-J_2)I/3}\,e^{i(\varphi_1+\varphi_2)/3-i\phi},\\
&\psi_3=\sqrt{(1+J_2)I/3}\,e^{i(\varphi_1-2\varphi_2)/3-i\phi}.
\end{split}\end{equation}
The rescaled Hamiltonian for the degrees of freedom
$(J_1,\varphi_1,J_2,\varphi_2)$ is given by
\begin{equation}\begin{split}
\frac{H-\bar\omega{I}}{(I/2\bar\omega)^2}={}&{}
(2\Omega+\Omega'-2/3)J_1+(\Omega+2\Omega'-2/3)J_2\\
{}&{}-\frac{2}3\sqrt{(1+J_1)(1-J_1-J_2)}\,(2-J_2)\cos\varphi_1\\
{}&{}+\frac{1}3\,(1+J_1)(1-J_1-J_2)\cos{2}\varphi_1\\
{}&{}-\frac{2}3\sqrt{(1-J_2)(1-J_1+J_2)}\,(2-J_1)\cos\varphi_2\\
{}&{}+\frac{1}3\,(1-J_2)(1-J_1+J_2)\cos{2}\varphi_2\\
{}&{}+\frac{7}3-\frac{2}{3}\,J_1J_2.
\end{split}\end{equation}
Still, due to the presence of trigonometric functions, the
numerical integration of the equations of motion for this
Hamiltonian would be less efficient than for the
polynomial Hamiltonian~(\ref{Htriple=}) with three degrees
of freedom.

\section{Numerical criterion for chaos}
\label{app:numerics}

\begin{figure*}
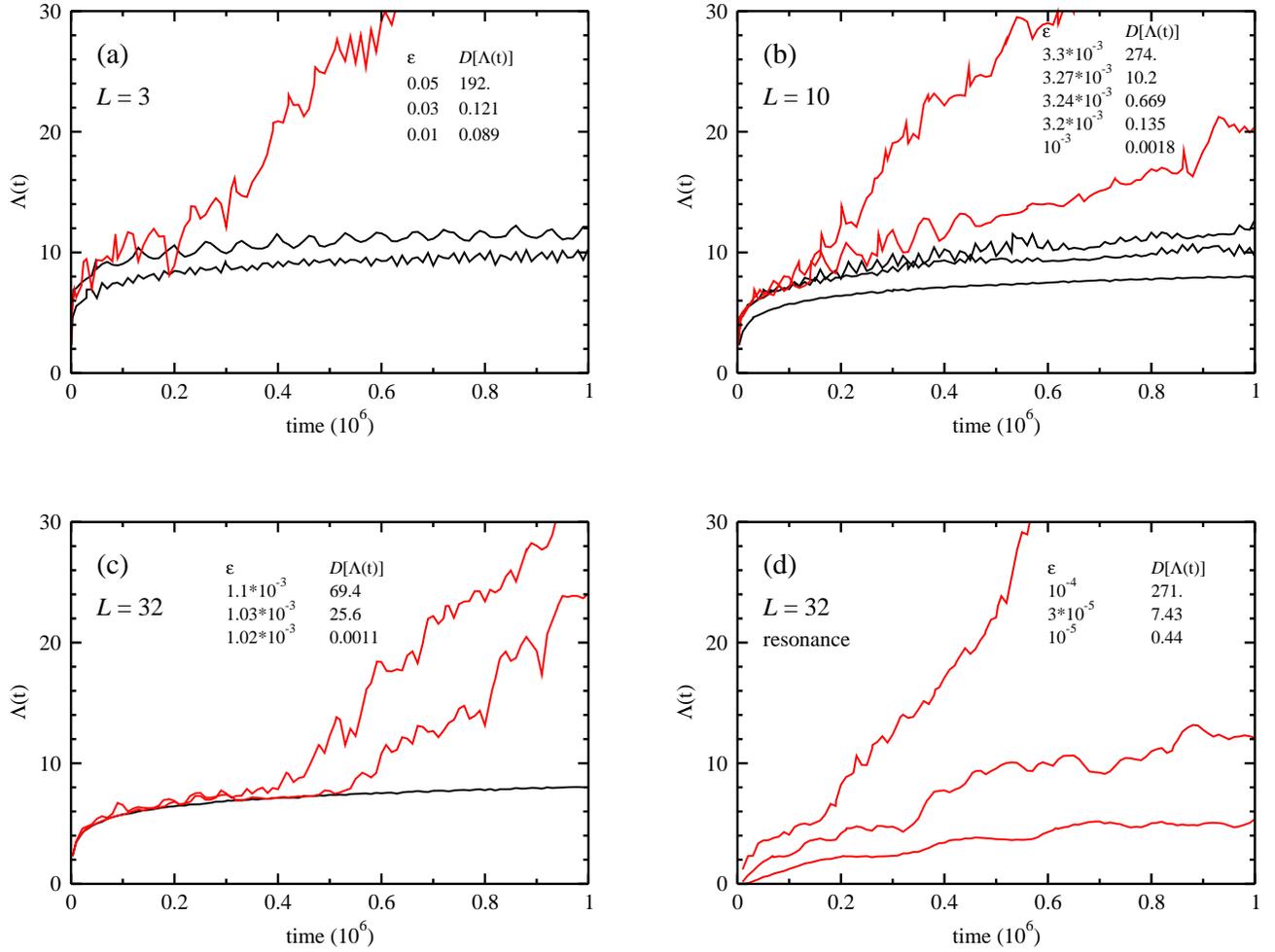

\includegraphics[width=8cm]{traj3}\hspace{1cm}
\includegraphics[width=8cm]{traj10}\vspace{1cm}
\includegraphics[width=8cm]{traj32}\hspace{1cm}
\includegraphics[width=8cm]{traj32spot}
\caption{\label{fig:traj}(color online)
$\Lambda(t)$ for $0<t<10^6$. Panels (a), (b), (c) correspond
to three chain lengths: (a)~$L=3$, (b)~$L=10$, (c)~$L=32$.
All curves on a given panel correspond to the same realization
of the disorder $\{\omega_n^2\}$, phases $\{\phi_n\}$, and
differ only by the value of $\ep_n=\ep$.
Higher curves correspond to larger values of~$\ep$.
Panel~(d) corresponds to $L=32$, but the disorder realization
was artificially modified by setting
$\omega_{12}=\omega_{13}=\omega_{14}$, thereby creating two
neighboring resonances.
}
\end{figure*}

The numerical criterion for the chaotic motion used in the
present work is based on the analysis of
$\Lambda(t)=\ln\frac{d(t)}{d(0)}$,
where $d(t)$ is the distance between points belonging
to two initially close trajectories.
To illustrate the idea, we plot several
traces $\Lambda(t)$ in Fig.~\ref{fig:traj}. Looking at them,
it is quite easy to guess which curve corresponds to a
regular motion, and which one to chaotic. The only exception
is the lowest curve in Fig.~\ref{fig:traj}(d). We know it
should be chaotic, as the corresponding realization of
$\{\omega_n^2\}$ contains two neighboring resonances,
introduced ``by hands''. However, at $\ep=10^{-5}$ energy
exchange between the oscillators is so slow, that following
their dynamics up to $T=10^6$ is insufficient to distinguish
between regular and chaotic motion. This gives us the lower
boundary on~$\ep$. Compare also the lowest curve in
Fig.~\ref{fig:traj}(c) and the middle curve in
Fig.~\ref{fig:traj}(d). The former one is regular and the
latter is chaotic, even though they have close values of
$\Lambda(t=T)$.
Thus, the fixed cutoff for the Lyapunov exponent (independent
of $L$~and~$\ep$), which was used in Ref.~\cite{Pikovsky2011},
as the criterion for chaos, may introduce systematic
errors and affect the scaling.

The main observation from Fig.~\ref{fig:traj} is that for
chaotic trajectories $\Lambda(t)$ grows linearly on the
average, corresponding to a finite Lyapunov exponent, while
for regular trajectories
\begin{equation}\label{Lambda=ln}
\Lambda(t)=\ln{t}+q_0,
\end{equation}
up to some weak noise, $q_0$~being some realization-dependent
constant.
This can be understood very simply: if two trajectories lie
on neighboring invariant tori, their frequencies are only
slightly different, so the phase mismatch between them is
accumulated slowly and linearly in time. The distance between
the trajectories is proportional to this phase mismatch, as
long as the latter is small compared to unity.

To determine how well a given function $\Lambda(t)$ is
approximated by the form~(\ref{Lambda=ln}) on an interval
$0<t<T$, consider the time average (denoted by overline):
$\overline{[\Lambda(t)-\ln{t}-q]^2}$.
It is a quadratic function of the parameter~$q$, which has
a minimum at some~$q$, and its value at the minimum is
$D\geq{0}$.
If $\Lambda(t)$ is given exactly by Eq.~(\ref{Lambda=ln}),
then simply $\overline{[\Lambda(t)-\ln{t}-q]^2}=(q-q_0)^2$,
so $D=0$. Thus, the value at the minimum, given by
\begin{equation}\label{DLambda=}
D[\Lambda(t)]=\overline{[\Lambda(t)-\ln{t}]^2}
-\left(\overline{[\Lambda(t)-\ln{t}]}\right)^2,
\end{equation}
measures the quality of the fit of $\Lambda(t)$ by the
expression~(\ref{Lambda=ln}). $D[\Lambda(t)]$~for all curves
in Fig.~\ref{fig:traj} is shown on the corresponding panel.
Thus, we consider the motion as regular if $D[\Lambda(t)]<1$
and chaotic otherwise.

When the Lyapunov exponent $\sigma>0$, it sets the natural
upper time limit for the numerical integration of the
differential equations. Indeed, when the double-precision
machine zero, $2^{-52}$, multiplied by $e^{\sigma{t}}$
becomes of the order of unity, the integration inevitably
deviates from the original trajectory, no matter how good
the integration scheme is. This corresponds to
$\Lambda=\ln{2}^{52}\approx{36}$. To check this, we have
performed the time-reversal test: at time $t=t_*$, we
invert all the momenta, and integrate up to $t=2t_*$.
Ideally, the system should return to its initial condition.
This was found to be the case if $\Lambda(t_*)<25-30$ (for
many trajectories whose Lyapunov exponents differed by one
or two orders of magnitude). Thus, considering larger
$\Lambda$ would produce the Lyapunov exponent not for a
given trajectory, but its certain average over the phase
space. As we are interested not in the trajectory itself,
but only in whether it is chaotic or regular, we consider
the very fact that $\Lambda(t)$ has reached 25 a sufficient
evidence for chaos. If this happens within the integration
time, $t<T$, we stop the integration, and simply set
$D[\Lambda]=30$.

\begin{figure}
\includegraphics[width=8cm]{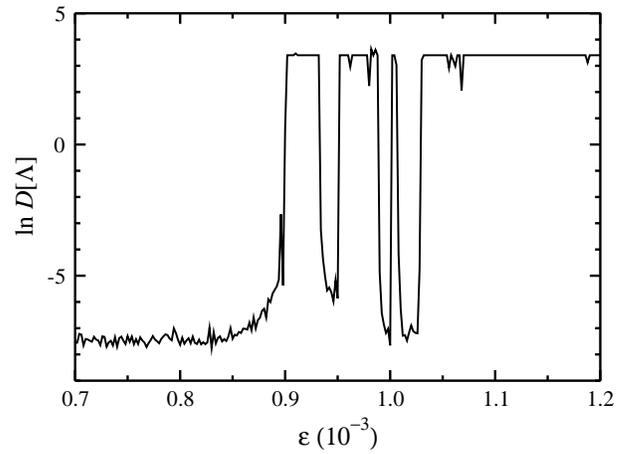}
\caption{\label{fig:reentrance}
$\ln{D}[\Lambda]$ [with $D$ defined in Eq.~(\ref{DLambda=})]
as a function of~$\ep$ for the same realization of disorder
and phases as in Fig.~\ref{fig:traj}(c). The upper cutoff
corresponds to $\ln{30}$, as discussed in the text.
}
\end{figure}

One could think that for every realization of the disorder
$\{\omega_n^2\}$ and of the oscillator phases $\{\phi_n\}$
exists a threshold value~$\ep_c$, such that for $\ep<\ep_c$
the motion is quasiperiodic, and for $\ep>\ep_c$ it is
chaotic. Indeed, a natural guess is that upon increase of
the integrability-breaking parameter~$\ep$ the chaotic
region of the phase space should grow. This would make the
calculation for the fixed-$\ep$ ensemble more efficient,
as one would not need to probe too small or too large
values of~$\ep$. However, this guess turns out to be wrong,
as seen from Fig.~\ref{fig:reentrance}, where we plot
$\ln{D}[\Lambda]$ as a function of~$\ep$ in the vicinity
of $\ep=10^{-3}$ for the same realization of disorder and
phases as in Fig.~\ref{fig:traj}(c). The flat regions on
the level of
$3.4\ldots=\ln{30}$ correspond to $\Lambda(t)$ reaching
$25$ for $t<T$, as discussed in the previous paragraph.
One can see that the conclusion about regular/chaotic
character of the dynamics is little sensitive to the
chosen numerical border $D=1$: had we chosen $D=5$ or
$D=1/5$, the result would not change. We also note that
for the particular realization, corresponding to
Fig.~\ref{fig:reentrance}, in all intervals of $\ep$
where the dynamics is chaotic, the Lyapunov eigenvector
is confined to sites from $n=21$ to $n=26$. This is
clearly related to the fact that in this realization
$|\omega_{22}-\omega_{24}|\approx{0.9}\cdot{10}^{-3}$,
so the observed reentrant behavior of chaos seems to
occur for the same guiding resonance.

\begin{figure}
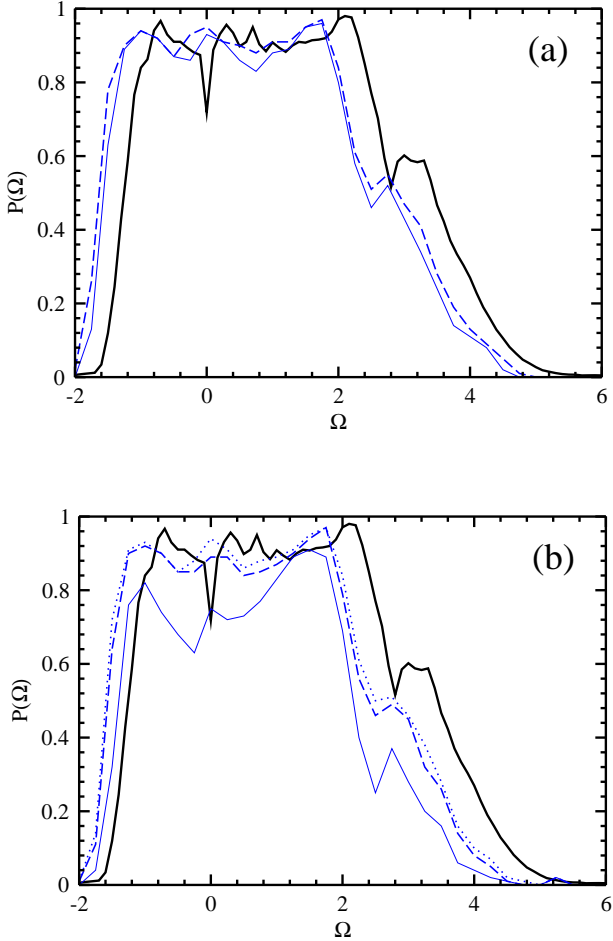

\vspace{0cm}
\includegraphics[width=8cm]{dissectionA}
\vspace{1cm}\\
\includegraphics[width=8cm]{dissectionB}
\caption{\label{fig:dissection}(color online)
The phase-averaged chaotic fraction $P(\Omega)$ for a
given realization of disorder on a chain with $L=32$
and frequencies $\omega_{12},\omega_{13},\omega_{14}$
chosen according to Eq.~(\ref{dissection=}),
for (a)~$\ep=10^{-4}$, (b)~$\ep=0.3\cdot{10}^{-4}$.
The thick solid line on both panels corresponds to
the calculation using the effective Hamiltonian from
Eq.~(\ref{Htriple=}) and represents the section of
Fig.~\ref{fig:fraction} along the line $\Omega'=0$.
The thin solid, dashed and dotted lines correspond to
$T=10^6,\,3\cdot{10}^6,\,5\cdot{10}^6$, respectively
[the last curve is present only on panel~(b)].
}
\end{figure}

To check how the results are affected by the choice
of the integration time~$T$, we take the same realization of
disorder as for the trajectories in Fig.~\ref{fig:traj} (the
chain length $L=32$), and set ``by hands'' three nearby
frequencies to be
\begin{equation}\label{dissection=}
\omega_{12}=\bar\omega+\frac{9}{4}\,\frac{\ep}{\bar\omega^3}\,\Omega,
\quad
\omega_{13}=\bar\omega-\frac{9}{4}\,\frac{\ep}{\bar\omega^3}\,\Omega,
\quad
\omega_{14}=\bar\omega,
\end{equation}
with a fixed $\bar\omega=\sqrt{0.7}$, two values of
$\ep=10^{-4},\,0.3\cdot{10}^{-4}$, and $\Omega$ varying in
the interval $-2<\Omega<6$. This chain is supposed to be well
described by the effective Hamiltonian in Eq.~(\ref{Htriple=})
with $\Omega'=0$. On the two panels of Fig.~\ref{fig:dissection}
we plot the phase-averaged chaotic fraction as a function
of~$\Omega$ for the two above-mentioned values of~$\ep$,
and for $T=10^6,\,3\cdot{10}^6,\,5\cdot{10}^6$, together with
the result of the calculation using Eq.~(\ref{Htriple=}) (the
section of Fig.~\ref{fig:fraction} along the line $\Omega'=0$).
For both values of~$\ep$, as $T$ is increased, the curves
approach the result of Eq.~(\ref{Htriple=}), up to an overall
horizontal shift. This horizontal shift appears because
Eq.~(\ref{dissection=}) neglects the nonlinear frequency shift
for the oscillators with $n=12$ and 14 due to their coupling to
the oscillators with $n=11$ and $15$, respectively [the
difference between primed and unprimed frequencies in
Eq.~(\ref{omegaprime=})]. This is also the reason why the dip
at $\Omega=0$ for the result of Eq.~(\ref{Htriple=}) is not
resolved on the curves for the long chain. What is important,
is that while at $\ep=10^{-4}$ the integration time $T=10^6$
is sufficient, for $\ep=0.3\cdot{10}^{-4}$ integration up to
$T=10^6$ clearly underestimates the chaotic fraction, so that
$T=3\cdot{10}^6$ is required.

Finally, we mention that the numerical integration of the
differential equations was performed using
(i)~the fourth-order Runge-Kutta method for the calculations
reported in this section, and
(ii)~Bulirsch-Stoer method with polynomial extrapolation for
the accumulation of statistics.
The latter method turns out to require about ten times less
CPU time that the former to reach the same level of accuracy.
The energy conservation was satisfied extremely well in all
calculated trajectories, including those with largest
Lyapunov exponents, even at times when the time-reversal test
had failed. Thus, it is unlikely that use of an integration
scheme which automatically respects some conservation laws of
the original equations (e.~g., a symplectic integrator which
preserves the phase space volume) would lead to more accurate
results for a given trajectory: the accuracy is lost primarily
in the directions, orthogonal to conservation laws.


\section{Single-site excitation}
\label{app:SingleSite}

\begin{figure*}
\includegraphics[width=8cm]{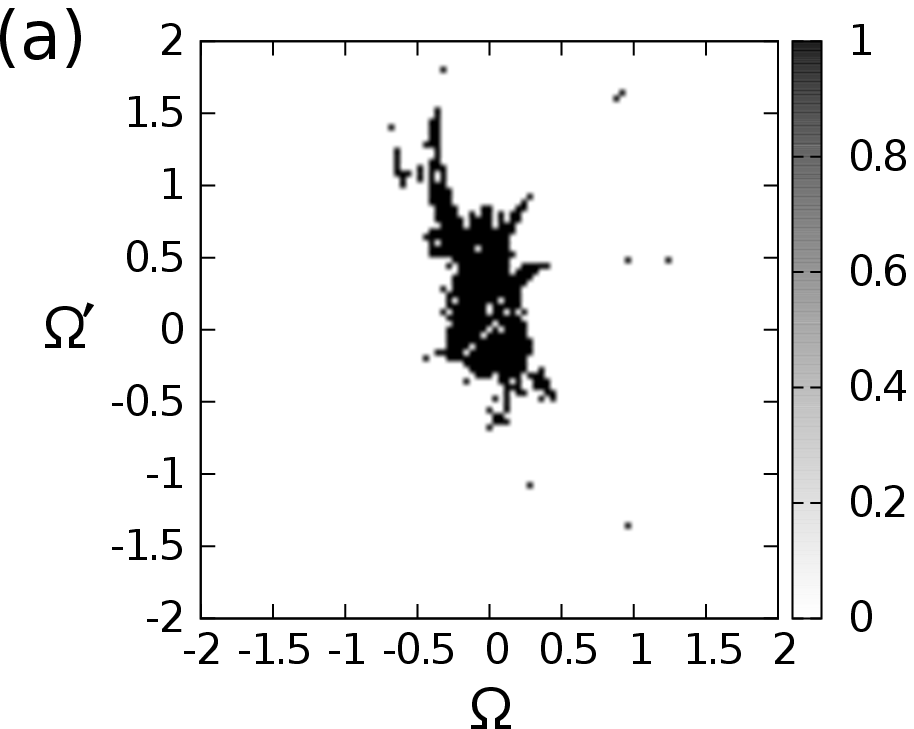}\hspace{1cm}
\includegraphics[width=8cm]{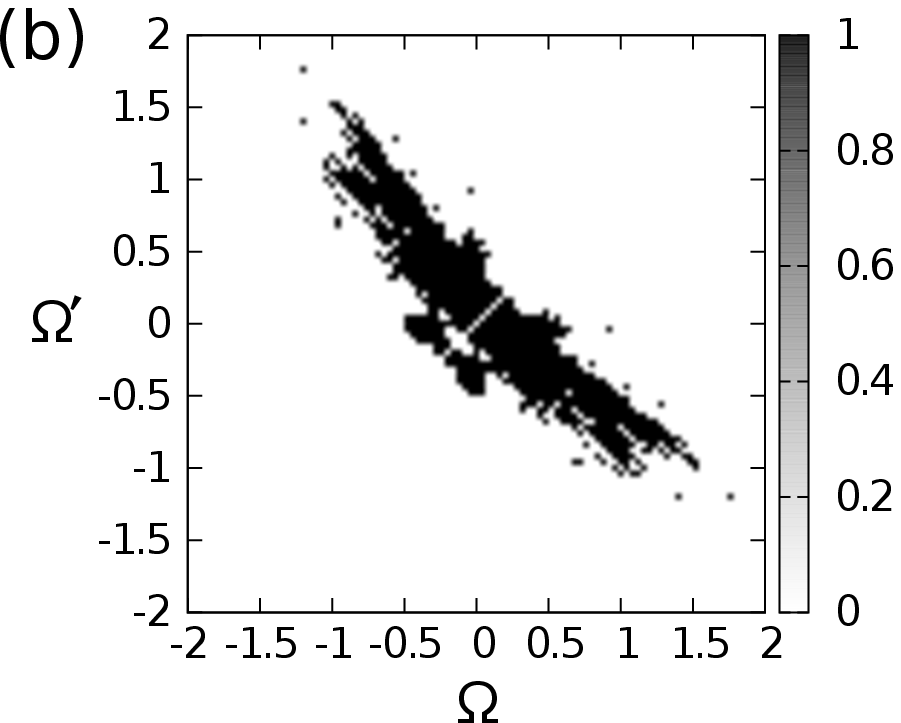}
\caption{\label{fig:SingleSiteTriple}
Values of $\Omega,\Omega'$ for which the single-site initial
conditions for the resonant triple Hamiltonian,
Eq.~(\ref{Htriple=}), lie on a chaotic trajectory (black spots)
and on a regular trajectory (white areas):
(a)~and~(b) represent the functions $\Theta_1(\Omega,\Omega')$
and $\Theta_2(\Omega,\Omega')$, corresponding to the initial
conditions (\ref{inicond1=}),~(\ref{inicond2=}), respectively.
}
\end{figure*}

Let us determine the probability of chaos for an initial
condition, localized on a single site,
\begin{equation}
p_n(0)=0,\quad q_n(0)=\delta_{n,(L+1)/2}\sqrt{2\ep}/\omega_n,
\end{equation}
for $L$~odd.
Chaos should occur if the resonant triple contains the site
$n=(L+1)/2$. Thus, the probability of chaos $P_1(\ep,L)$
should not depend on $L$ and scale as $P_1(\ep)=A_1\ep^2$,
at large~$L$ and small~$\ep$.

To determine the coefficient~$A_1$, we again use the effective
Hamiltonian of the triple, Eq.~(\ref{Htriple=}), and consider
two initial conditions:
\begin{subequations}\begin{eqnarray}
&&\Psi_1(0)=\frac{1}{\sqrt{3}},\quad
\Psi_2(0)=\Psi_3(0)=0,\label{inicond1=}\\
&&\Psi_1(0)=0,\quad\Psi_2(0)=\frac{1}{\sqrt{3}},\quad
\Psi_3(0)=0.\label{inicond2=}
\end{eqnarray}\end{subequations}
i.~e., the excitation to be initially localized on one
of the lateral sites of the triple, or on the central site.
Let $\Theta_{1,2}(\Omega,\Omega')$ equal one if the
corresponding trajectory is chaotic and zero if it is regular,
for each of the two initial conditions, respectively.
The functions $\Theta_{1,2}(\Omega,\Omega')$ are shown in
Fig.~\ref{fig:SingleSiteTriple}.
The chaotic probability is then given by
\begin{equation}
P_1(\ep)=81\ep^2\int\limits_{-\infty}^\infty d\Omega\,d\Omega'
\left[2\,\Theta_1(\Omega,\Omega')
+\Theta_2(\Omega,\Omega')\right].
\end{equation}
For the initial condition~(\ref{inicond1=}), the
$\Omega,\Omega'$-integral is equal to $0.75\pm{0}.02$,
while for Eq.~(\ref{inicond2=}) it is equal to
$1.39\pm{0.02}$,
which gives $A_1\approx{235}$.


\begin{thebibliography}{99}
\bibitem{Eperosc}
Strictly speaking, $\ep_n$~does not correspond to the total
energy per oscillator, as it does not include~$H_\mathrm{int}$,
the second term of Eq.~(\ref{HFSW=}). In particular, this means
that (i)~the sum of all $\ep_n$, corresponding to~$H_0$ is
not conserved during the evolution, and (ii)~Eq.~(\ref{Pfix=})
with the $\delta$-function replaced by the exponential is not
exactly the thermal distribution.
However, $H_0\sim{L\ep}$ and $H_\mathrm{int}\sim{L\ep^2}$, so
$H_\mathrm{int}$ represents a small correction at $\ep\to{0}$,
and $\ep$ indeed has the meaning of temperature in the
thermal ensemble.
\bibitem{chaosDNLS}
This is because the system~(\ref{HFSW=}) has only one integral
of motion (energy), which should be contrasted with DNLS which
has two integrals of motion (energy and norm), so the minimal
number of oscillators needed to produce chaos in DNLS is three.
\end{thebibliography}
\end{document}